\documentclass[conference]{IEEEtran}

\AtBeginDocument{%
  \providecommand\BibTeX{{%
    \normalfont B\kern-0.5em{\scshape i\kern-0.25em b}\kern-0.8em\TeX}}}

\usepackage[frozencache,cachedir=.]{minted}

\usepackage{booktabs}
\let\cline\cmidrule
\usepackage{cite}
\usepackage{amsmath,amsfonts}
\usepackage{algorithmic}
\usepackage{graphicx}
\usepackage{textcomp}
\usepackage{xcolor}
\usepackage{minted}
\usepackage{hyperref}
\usepackage{multirow}
\usepackage{enumitem}           
\usepackage[T1]{fontenc} 
\usepackage{tikz} 
\usepackage{stfloats}
\usepackage[utf8]{inputenc}
\usepackage{soulutf8}
\usepackage{balance}
\usepackage[caption=false]{subfig} 

\newcommand{\figref}[1]{Fig.~\ref{#1}}

\newcommand{\listref}[1]{Listing~\ref{#1}}

\newcommand{\ie}{\textit{i.e.},\ }
\newcommand{\eg}{\textit{e.g.},\ }
\newcommand{\etal}{\textit{et al.} }
\newcommand{\etc}{{\em etc.}}

\definecolor{francBlue}{RGB}{64,76,87}
\newcommand\circled[1]{%
  \tikz[baseline=(X.base)] 
    \node (X) [line width=0pt,draw, shape=circle, inner sep=0, fill=francBlue, text=white] {\strut  \bf #1};%
}

\usepackage[most]{tcolorbox}
\usepackage[parfill]{parskip}

\tcbset{
  parskip/.style={
    before={\par\pagebreak[0]\parindent=0pt},
    after={\parfillskip=0pt\par},
  },
}
\newtcolorbox{resultbox}[1][]{%
    colback=blue!3,
    colframe=blue!3,
    notitle,
    sharp corners,
    borderline west={2pt}{0pt}{gray!80!black},
    enhanced,
    breakable,
    boxsep=0pt,
    left=4pt,right=2pt,top=2pt,bottom=2pt,
    }

\newcommand{\rques}[1]{  
\begin{tcolorbox}[enhanced jigsaw,colback=white,left=2pt,right=2pt,top=2pt,bottom=2pt]
#1
\end{tcolorbox}
}
\definecolor{codebg}{rgb}{0.99,0.99,0.99}
\definecolor{hiliteColor}{rgb}{1,0.92549019607,0.6}
\definecolor{tainted}{rgb}{0,1,1}

\newmintinline{python}{fontsize=\normalsize}

\newminted[PythonSourceCode]{python}{
    labelposition=topline,
    frame=single,
    numbersep=2pt,
    fontsize=\tiny,
    xleftmargin=5pt,
    framerule=0.5pt,
    linenos
}

\newminted[JavaSourceCode]{java}{
    labelposition=topline,
    frame=single,
    numbersep=2pt,
    fontsize=\tiny,
    xleftmargin=5pt,
    framerule=0.5pt,
    linenos
}

\definecolor{magnolia}{rgb}{0.97, 0.96, 1.0}
\definecolor{shadecolor}{rgb}{0.97, 0.96, 1.0}
\newcommand{\code}[1]{\texttt{\small{#1}\normalsize}}
\newmintinline[codePython]{python}{fontsize=\small}
\newmintinline[codeJava]{java}{fontsize=\small}
\newmintinline[snippetJava]{java}{fontsize=\small,bgcolor=magnolia}
\newmintinline[snippetPython]{python}{fontsize=\small,bgcolor=magnolia}
\newmintinline[snippetPerl]{perl}{fontsize=\small,bgcolor=magnolia}

\newcommand{\franc}[0]{\textsc{Franc}}

\newcommand{\hlfancy}[1]{\sethlcolor{magnolia}\texttt{\small\hl{#1}}\sethlcolor{yellow}}

\begin{document}

\title{\franc: A Lightweight Framework for High-Quality Code Generation}

\author{
\IEEEauthorblockN{Mohammed Latif Siddiq}
\IEEEauthorblockA{\textit{University of Notre Dame} \\
Notre Dame, IN. USA \\
msiddiq3@nd.edu}
\and
\IEEEauthorblockN{Beatrice Casey}
\IEEEauthorblockA{\textit{University of Notre Dame} \\
Notre Dame, IN. USA \\
bcasey6@nd.edu}
\and
\IEEEauthorblockN{Joanna C. S. Santos}
\IEEEauthorblockA{\textit{University of Notre Dame} \\
Notre Dame, IN. USA \\
joannacss@nd.edu}
}

\maketitle

\begin{abstract}
In recent years, the use of automated source code generation utilizing transformer-based generative models has grown in popularity. These models can generate code according to the developers' requirements. However, recent research showed that these automatically generated source codes can contain vulnerabilities and other quality issues. Despite researchers' and practitioners' attempts to enhance code generation models, retraining and fine-tuning large language models is not only time-consuming but also resource-intensive and costly. Thus, in this paper, we describe \franc, a lightweight framework for recommending more secure and high-quality source code derived from transformer-based code generation models. 
\franc{} includes a static filter to make the generated code compilable with heuristics and a quality-aware ranker to sort the code snippets based on a quality score. Moreover, the framework uses prompt engineering to fix persistent quality issues. 
We evaluated \franc{} with five Python and Java code generation models and six prompt datasets, including a newly created one in this work (\franc{}). The static filter improves 9\% to 46\% Java suggestions and 10\% to 43\% Python suggestions regarding compilability. The average improvement over the NDCG@10 score for the ranking system is 0.0763, and the repairing techniques repair the highest 80\% of prompts.  \franc~takes, on average, 1.98 seconds for Java; for Python, it takes 0.08 seconds.
\end{abstract}



\section{Introduction}\label{sec:intro}

Large Language Models (LLMs) are increasingly performing well in many natural language processing tasks, such as text generation, translation, and summarization~\cite{zan2023nl2code}. One particularly relevant application of these LLMs to software engineering is \textit{code generation}, in which LLMs are trained and fine-tuned with \textit{large} amounts of \textit{code} snippets~\cite{chen2021codex,Aman22, codet,coderl2022}. 
These techniques help developers by    generating the implementation of functions/programs based on provided \textit{prompts}~\cite{Le_2021}.  




Although LLM-based code generation techniques may produce \textit{functionally} correct code, they can contain \textit{code smells} and \textit{vulnerabilities}~\cite{siddiq2022empirical, pearce2021,sandoval2022security}. A recent study showed that LLMs are fine-tuned with samples containing harmful coding patterns that leak to the generated code~\cite{siddiq2022empirical}. Another study found that GitHub Copilot can produce vulnerable code~\cite{pearce2021}. With the increasing use of LLM-based code assistants, these problematic code snippets can get deployed into production, negatively affecting the software system's security and reliability.


To improve the quality of the generated code, we first need a \textit{good dataset} (\ie free of quality issues)~\cite{gunasekar2023textbooks}. However, collecting training data is \textbf{\textit{time-consuming}}~\cite{Najafabadi2015} and \textbf{\textit{challenging}} because  code datasets commonly used for training contain \textbf{\textit{quality issues}} (\eg bugs, vulnerabilities, and code smells~\cite{sharma2017house,palomba2018diffuseness,siddiq2022empirical}) which would require vetting these code samples to remove or repair low quality training samples. Moreover, fine-tuning an LLM model is a \textbf{\textit{resource-hungry}} process~\cite{church_chen_ma_2021} which requires \textit{at least one} GPU, and pre-training is typically performed on a large cluster of GPUs.  
Although models can make inferences without GPUs, the throughput may not be optimal. For example, we used a GPU with 24 GB RAM when using a model with 2.7 billion parameters to generate code  with 128 tokens (and up to 2,048 context tokens). To generate more tokens and/or larger models, more GPUs or a more expensive GPU with a larger RAM would be needed. 

Although LLM fine-tuning can be done in the cloud to avoid acquiring expensive GPUs, it is still costly.
For example, OpenAI's fine-tuning API currently costs U\$0.03 dollars per 1,000 tokens (8K context GPT-4 model)~\cite{openaipricing}. Since a large dataset may have billions of tokens, fine-tuning this model would cost thousands of dollars. Besides \textit{fine-tuning} costs, there is a separate cost for \textit{inference}. It costs U\$0.06 per 1,000 tokens to use your own fine-tuned GPT-4 model. 

While recent works~\cite{he2023large,chen2023improving,Ding24cycle} aimed to improve the output of generated code, they focus on a specific quality attribute (\eg security) and require  collecting training samples to train/fine-tune the models.
Since LLMs can generate code with \textit{quality issues}, we need a \textbf{non-expensive way} to provide the \textit{best} generated code to the user. 
In light of this need, this paper describes \franc, a \textbf{\textit{lightweight}, \textit{configurable}, and \textit{model-agnostic} framework to \textit{filter}, \textit{rank}, and \textit{repair} code automatically generated by LLMs}. 
\franc{} (Filter, Rank, And model-agNostic Configurable) framework works by taking as input a developer's prompt and then using \textbf{(i)} static filters to remove \& repair non-compilable code, \textbf{(ii)} off-the-shelf quality issue detection tools to rank generated code snippets with respect to their measured quality, and \textbf{(iii)} generated knowledge prompting to fix quality issues. 

To demonstrate its effectiveness, we conducted an empirical evaluation in which we used \franc{} to improve the quality of Java and Python code generated by five models (CodeParrot, InCoder, CodeGen, PolyCoder, and GPT-3.5-turbo).  In this experiment, we generated code from \textbf{1,081}   prompts collected from existing code generation benchmarks~\cite{hao2022aixbench,yu2023codereval,2023multilingual,chen2021codex,siddiq2022seceval}  and \textbf{70} prompts we created from questions posted on StackOverflow. This framework also can effectively repair suggestions to ensure high-quality code in the output.  

This paper's \textbf{contributions} are \textbf{\textsf{(1)}} a novel lightweight framework (\franc{}) to filter, rank and repair the output of code generation models based on code quality; \textbf{\textsf{(2)}} automated filtering capabilities to remove non-compilable and unnecessary portions of the generated code to minimize human inspection; \textbf{\textsf{(3)}} a demonstration of how prompt engineering (and different prompt repair structures) can help to repair quality issues automatically; \textbf{\textsf{(4)}} an empirical investigation comparing the effectiveness of the framework with existing code generation and infilling models; \textbf{\textsf{(5)}} a dataset of \textbf{70} prompts. This paper's replication package is available at: \url{https://github.com/s2e-lab/FRANC}.

\section{Background}
\label{sec:background}
This section explains key concepts used in this work.
\subsection{LLM-based Code Generation}
\textbf{\textit{LLM-based code generation}} leverages NLP techniques to generate source code from a given \textbf{\textit{prompt}}. The prompt can be a combination of natural language and  code (\eg a function definition, expressions, code comments \etc). The code generation problem was previously tackled as a \textit{sequence-to-sequence (seq2seq) learning} problem~\cite{seq2seq}. Prior works used Recurrent Neural Networks (RNN) and neural networks based on Long Short-Term Memory (LSTM)~\cite{seq2seq, Sherstinsky_2020} to generate source code. The LSTM and RNN use a feedback loop during training to memorize specific portions of the input for learning.

The attention-based transformer architecture revolutionized the field of language learning in 2017. The transformer is based on an encoder-decoder architecture that leverages the self-attention mechanism to weigh the importance of each input data point \cite{attention2017}. There are several transformer-based deep learning models, like \textbf{\texttt{BERT}} (Bidirectional Encoder Representations from Transformers) \cite{bert2018},
 \textbf{\texttt{T5}} (Text-to-Text Transformer) \cite{2020t5} and \textbf{\texttt{GPT-3}} (Generative Pre-trained Transformer) \cite{brown20}. 
These language learning models can be fine-tuned with code-related datasets for software engineering tasks such as source code \textit{completion} \cite{izadi2022codefill,kim2021code,svyatkovskiy2021fast}, \textit{search} \cite{codebert}, and \textit{summarization} \cite{gao2022m2ts}. Examples of this type of model include \textbf{\texttt{CodeBERT}} \cite{codebert}, \textbf{\texttt{CodeT5}} \cite{codet5}, and \textbf{\texttt{Codex}} \cite{chen2021codex}.

\subsection{Code Quality and Code Smells}
\textbf{\textit{Code quality}} is a broad term used to refer to code snippets that are free of bugs and conformant to its requirements~\cite{QualityBook}. The quality of a source code is usually expressed in terms of  \textit{defect rate} (\ie measured as the number of defects per unit, \eg lines of code, function points, \etc) and \textit{reliability}  (measured in terms of failure occurrences, \eg the number of failures during a period, mean time to failure, \etc). However, the main indicator of high-quality code that complies with specifications depends on the vendors' demands. 
To ensure code quality, a common \textbf{\textit{coding standard}} needs to be adopted by every contributor. Different languages adopt specific coding practice protocols. For example, PEP-8~\cite{pep8} is a well-known guide for coding practice standards for Python. It provides an extensive guide for code layout, whitespace usage, and naming conventions \etc~For instance, according to the guide, the coding layout should have 4 spaces per indentation level, and spaces are the preferred indentation method.

An indication of poor system design and implementation practices is a \textbf{\textit{code smell}} (also known as a ``bad code smell'' or ``smell'')~\cite{fowler1999}. These code smells can introduce software maintenance problems. Furthermore, they contravene fundamental software design rules, reducing the product's potential effectiveness. These issues increase the possibility of future errors or failures or can hinder software development \cite{pereira22}, affecting the software's reliability. For example, the code in Listing~\ref{lst:smell} may throw a \code{ValueError} if the input at line  3 cannot be parsed to \codePython{int}. Although there is a \code{try-except} block to catch the exception (highlighted), the exception is not handled, which is an example of a code smell \cite{gupta2018software}.

\begin{listing}[!ht]
\caption{Examples of a code smell and security smell}\label{lst:smell}
\noindent\begin{minipage}{.495\linewidth}    
\begin{PythonSourceCode*}{highlightlines=4-5,label=\tiny{\textsf{Code smell example}}}
try:
    num = input('Enter number:')
    num = int(num)
except ValueError:
    pass
\end{PythonSourceCode*}
\end{minipage}\hfill
  \begin{minipage}{.495\linewidth}
    \begin{PythonSourceCode*}{highlightlines=3-3,label=\tiny{\textsf{Security smell example}}}
import hashlib
def validate(c, h):
    hash_md5 = hashlib.md5(c)
    hash = hash_md5.hexdigest()
    return hash == h
\end{PythonSourceCode*}
  \end{minipage}
\end{listing}

Insecure coding, flaws with design choices, and coding standard violations all fall under the umbrella phrase ``{code smell}''. \textbf{\textit{Security code smells}} (or simply ``security smells'') are a \textit{subset} of code smells. They are frequently used programming patterns that could result in vulnerabilities~\cite{rahman_seven_2019,rahman2019share}. Security smells point to the possibility of a vulnerability, even if they may not constitute vulnerabilities entirely by themselves~\cite{ghafari2017security}. For example, the code (shown on the right) in Listing~\ref{lst:smell} is taken from CodeQL examples \cite{codeql}. It uses the \texttt{md5} hash function that is unsafe, which is related to \textit{CWE-327: Use of a Broken or Risky Cryptographic Algorithm}~\cite{siddiq2022empirical}.

In our work, we focus on the \textbf{\textit{quality of automatically generated code}} with respect to following \textit{code standards} and the absence of \textit{code smells} and \textit{security smells}. Specifically, our framework aims to ensure that the \textit{top generated code} is free of quality issues, as engineers  often focus on the first generated code~\cite{roy2022systematic,wu2024repoformer}.

\subsection{Motivating Example}\label{subsec:MotivatingExample}
A code generation model can generate multiple (\textit{ranked}) suggestions for a given prompt. However, some produced suggestions may contain \textit{quality issues} (\eg security smells)~\cite{siddiq2022empirical}. For example, consider that we provide the prompt in Listing~\ref{lst:copilotVulnerability} (highlighted lines) to GitHub Copilot \cite{copilot}. Although the   generated code (lines 8-11) on the first position of GitHub Copilot's rank is \textit{functionally correct}, it contains a {\textit{SQL injection vulnerability}} because it uses a formatted string to construct the query (line 9). 
\begin{listing}[!ht]
\caption{Example the top-1 code generated by GitHub copilot}\label{lst:copilotVulnerability}
\includegraphics[width=\linewidth]{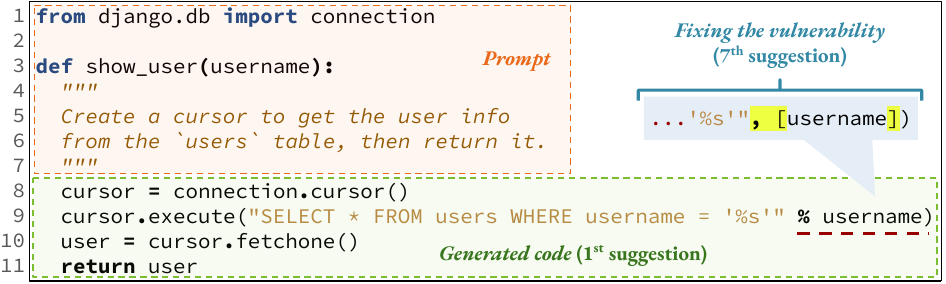}
\end{listing}


When we inspect the remaining generated suggestions, we can see code snippets that are still functionally correct but that \textit{do not} contain a SQL Injection (or any other vulnerability). For instance, one of the lowest-ranked suggestions (shown as a callout Listing~\ref{lst:copilotVulnerability}) uses a parameterized query to avoid SQL injection vulnerability. Since this suggestion is not the top one recommended generated by the model as developers are biased towards the first suggestion they see \cite{barke2023grounded}, we need a lightweight ranking system to address the quality issues of the generated suggestions. Thus, our goal is \textbf{to show as the \emph{top suggestion} to the user a source code that is vulnerability-free, smell-free code, and follows standard coding practices.}



\section{Our Framework (\franc)}\label{sec:method}

\begin{figure*}[!ht] 
    \centering
\includegraphics[width=\textwidth]{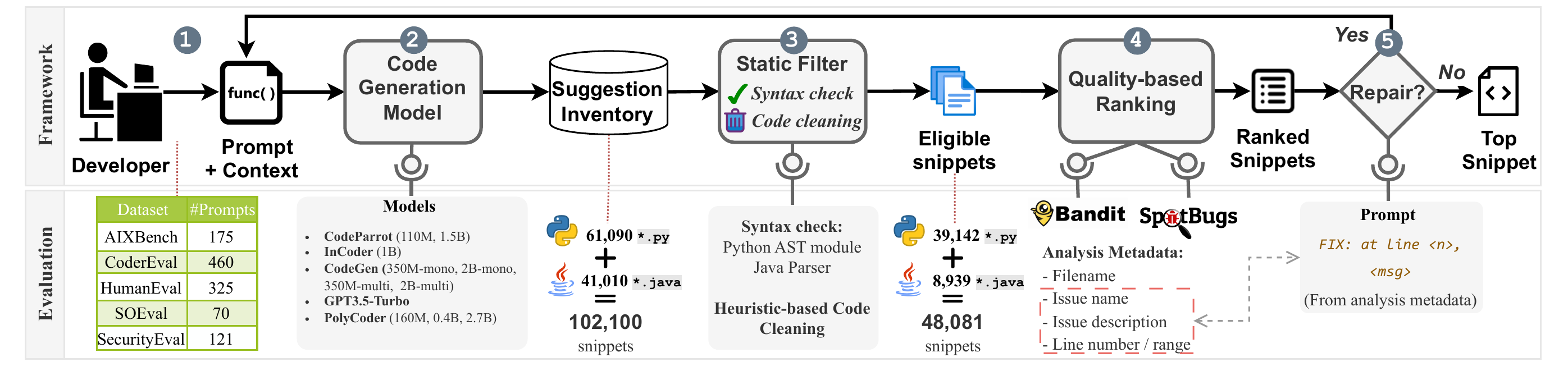}    \caption{Framework overview and evaluation parameters} \label{fig:method}
\end{figure*}  

\figref{fig:method} provides an overview of our lightweight quality-aware framework (\franc) for code generation (\textit{``Framework''} swimlane). \franc{} has five phases that are \textit{\textbf{model-agnostic}}. Each phase can be tailored to the underlying programming language and LLM being used by the developer (as shown in the \textit{``Evaluation''} swimlane). 

During the \circled{1} \textit{\textbf{\underline{prompt and context creation}}} phase, software engineers specify the code's expected  behavior.
\circled{2} In the \textit{\textbf{\underline{code generation}}} phase, \franc{} uses an existing LLM to generate code snippets. Since these models may generate an $n$ amount of sorted samples, the output of this phase is the framework's \textit{suggestion inventory} (\ie a sorted set of $n$ automatically generated code snippets).
\circled{3} In the lightweight \textit{\textbf{\underline{static filtering}}} phase, \franc{} applies rules to generated code in order to automatically fix and/or remove code snippets in the suggestion inventory with syntax errors.  The output of this phase is a set of $x$ \textit{eligible code snippets} that passed the filtering criteria ($x \leq n$).
\circled{4} In the \textit{\textbf{\underline{quality-based ranking}}} phase, \franc{} sorts the code snippets from the previous phase based on a configurable \textit{quality score}. 
\circled{5} In the \textit{\textbf{\underline{repairing}}} phase, \franc{}  repairs problematic generated code via \textit{generated knowledge prompting}~\cite{liu2022generated} in which \franc{} crafts a prompt that instructs the code generation model to fix quality issues (\eg \textit{``Fix the buffer overflow at line 10''}).

Our  framework relies on two insights. \textbf{\textit{\underline{Insight 1}}}: as shown in Section~\ref{subsec:MotivatingExample}, although the \textit{first} generated code by a LLM may be insecure, the model can also generate alternative versions that do not contain a vulnerability, but that are ranked at lower positions. For instance, the correct (and secure) code shown in Listing~\ref{lst:copilotVulnerability} was also generated by GitHub Copilot for the given prompt, but it was ranked on the 7\textsuperscript{th} position. \textbf{\textit{\underline{Insight 2}}}: code repair can be treated as a code generation task \cite{Pearce2023,joshi2022}, but with a prompt that explicitly instructs the model to fix a quality problem in a given location. 
In the next sections, we detail each of the five phases of our framework.

\subsection{Phase 1: Prompt Creation}\label{subsec:Phase1}

First, the software engineer creates a \textit{prompt} for the code generation model.
The lines 1--6 (highlighted in blue) in the code snippet in~\listref{lst:gen_code} is an example of input an engineer can provide to an LLM. It contains an import statement, a function declaration, and a comment describing the function's intended functionality. 
This prompt instructs the LLM to generate the body of the function \codePython{yaml_load(filename)} that will parse a YAML file, create an object with the loaded data, and return it to the caller of the function.

\subsection{Phase 2: Code Generation}
In the second phase, \franc{} uses an LLM to generate code. Since our framework is model-agnostic, the model can be an open-source model (\ie datasets and parameters are publicly available) or a closed-source one (commercial). 
The model takes as input the \textit{prompt} and produces multiple sorted code suggestions (code snippets). The model's sorted list of code snippets composes \franc's \textit{\textbf{suggestion inventory}}. For instance, the code in Listing~\ref{lst:gen_code} is the top suggestion generated by  CodeGen  (2 billion parameters)~\cite{Nijkamp2022ACP} when given the prompt described in Section~\ref{subsec:Phase1}. This code repeats the prompt (lines 1-6), provides an implementation to \codePython{yaml_load(filename)}, and contains an \textit{unnecessary} and  \textit{incomplete}  function (lines 10-18). This incomplete function  serializes objects into a YAML file, which does not match the intent in the prompt and has a syntax error (missing closing parenthesis at line 18). Thus, in the next phase, \franc{} removes this unnecessary function and other types of syntax issues.

\begin{listing}[!ht]
        
\includegraphics[width=\linewidth]{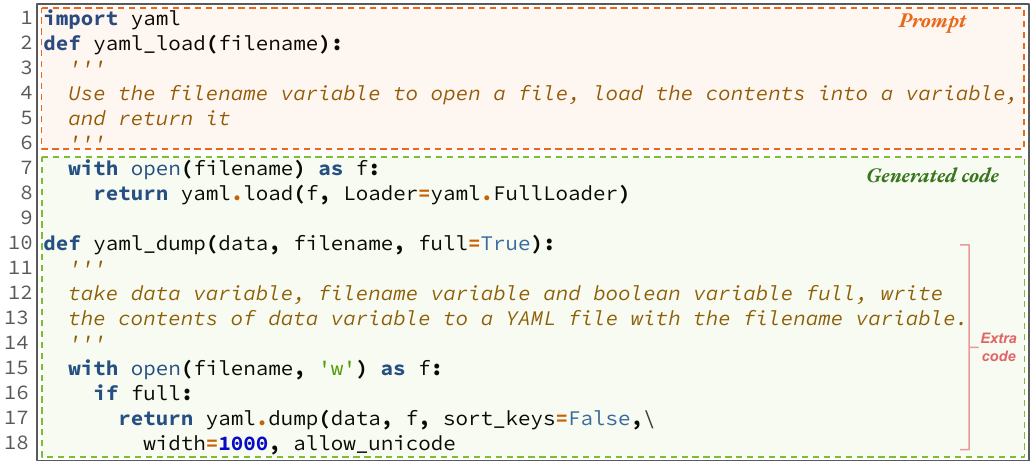}
\caption{Generated code example}\label{lst:gen_code}
\end{listing}

\subsection{Phase 3: Static Filtering}
\label{subsec:phase3}

As shown in \listref{lst:gen_code}, code snippets in the suggestion inventory may include \textit{unnecessary} blocks of code (\ie lines of code that do not match the intent specified in the prompt). Moreover, they may have \textit{syntax errors}~\cite{siddiq2022empirical}. Consequently, developers need to manually vet and fix the code when it is not compilable or contains unnecessary extra code. 
Therefore, in this third phase, our framework inspects each snippet from the \textit{Suggestion Inventory} and applies a set of \textbf{\textit{rules}} to \textit{(a)} remove unnecessary code from the snippet, and \textit{(b)} check whether it is syntactically correct. The output of this phase is a set of \textit{\textbf{eligible snippets}} that had any extra code removed from it \textbf{(a)} and passed the syntax check \textbf{(b)}.

The rules used for this filtering phase are \textit{configurable}, meaning that they are tailored to the underlying programming language being used.
For example, a rule that can be applied to clean the Python code in ~\listref{lst:gen_code} is to remove any code \textit{after} from the prompt's function. The resulting snippet would not contain lines 10-19.

\subsection{Phase 4: Quality-based Ranking}\label{sec:QualityRanking}

Automatically generated code can contain several quality issues, such as coding standard violations, code smells, and vulnerabilities~\cite{pearce2021,siddiq2022empirical,siddiq2022seceval}. 
Thus, the ranking used  by the code generation model may produce code with quality issues as the first (top-1) suggestion. 
Hence, \franc{} includes a \textit{configurable quality-based ranker} that sorts code snippets in a model's output based on a \textbf{\textit{quality score}} $\mathbf{Q(c_i)}$:

\vspace{-10pt}
\begin{equation}\label{eq:rank}
    Q(c_i) = \sum\limits_{j=1}^m w_j  q_j(c_i) \text{ where } \smash{\sum\limits_{j=1}^m w_j=1} \text{ and }  w_j \ge 0
\end{equation}
\vspace{-15pt}

Each \textit{code snippet} $c_i$ ranked at the position $i$ in the model's output is evaluated according to different \textit{quality factors} $q_j(c_i)$ that take into account a specific quality attribute (\eg security, performance, code smells \etc). Each \textit{quality factor} ($0 \le q_j(c_i) \le 1$) has a corresponding non-negative \textit{weight} $w_j$. Thus,
the quality score $Q(c_i)$ is a weighted average of each quality factor $q_j(c_i)$, ranging from \textbf{0} (\textit{lowest} quality) to \textbf{1} (\textit{highest} quality). 
\franc{} ranks all code snippets based on their quality score $Q(c_i)$ and presents the best one to the developer. If multiple snippets have the same score, \franc{} keeps the original order from the code generation model. For example, if the code snippets $c_3$ and $c_8$ have the same highest quality score, \ie $Q(c_3) = Q(c_8)$, then \franc{} chooses $c_3$ to be in the first position.

\subsection{Phase 5: LLM-based Code Quality Repair}\label{subsec:RepairPhase}
Although \franc{} has a quality-based ranking phase to ensure that the highest quality code snippet is given to the developer, it might be the case that the \textit{top 1} code suggestion includes quality issues because \textit{\textbf{all} code snippets generated by the model had quality issues}. Consequently, that would require the engineer to go through the burden of \textit{manually identifying} the quality problems in the generated code and \textit{fixing} them.
Hence, \franc{} includes a  \textit{configurable automated repairing} phase that relies on the LLM itself to fix the problematic lines in the code snippet. The key insight of this phase is that code repair can be treated as a code generation task \cite{Pearce2023,joshi2022}, but with a prompt that explicitly instructs the model to fix a quality problem in a given location.

\franc{} repairs code samples via \textit{Generated Knowledge Prompting} (\ie prompting for a task by incorporating knowledge or information \cite{liu2022generated}). In this case, \franc{} uses knowledge from the static analyzers to create the prompts. If the top-1 code snippet after the quality-based ranking has a quality score \textit{below an acceptable threshold} $\mathbf{\tau}$ (\ie $Q(c_1) < \mathbf{\tau}$), then \franc{} creates a \textit{\textbf{repair prompt}} and re-passes this engineered prompt to the model to fix the code. 


\franc{} uses the issue's description and location detected by a static analyzer to craft a \textbf{\textit{repair prompt}}  using three different structures (\textbf{P1}, \textbf{P2}, and \textbf{P3}) and send it back to the code generation model to generate code again. To illustrate these different prompt repair structures, consider the generated code to be repaired shown in Listing~\ref{lst:codeToRepair} (where the highlighted lines were the original prompt used to generate this code). The prompt structure \textbf{P1} adds source code comments \textit{after} the code to be repaired in the format ``\hlfancy{Fix: At line « \# », « error msg »\textbackslash{n} Fixed Code:\textbackslash{n}}'', as shown in Listing~\ref{lst:codeToRepair} (lines 10-11).  The error message comes from a static analyzer (in this example, it comes from Bandit).

\begin{listing}[!ht]
\caption{Generated code with a SQL injection (line 7)}\label{lst:codeToRepair}
\includegraphics[width=\linewidth]{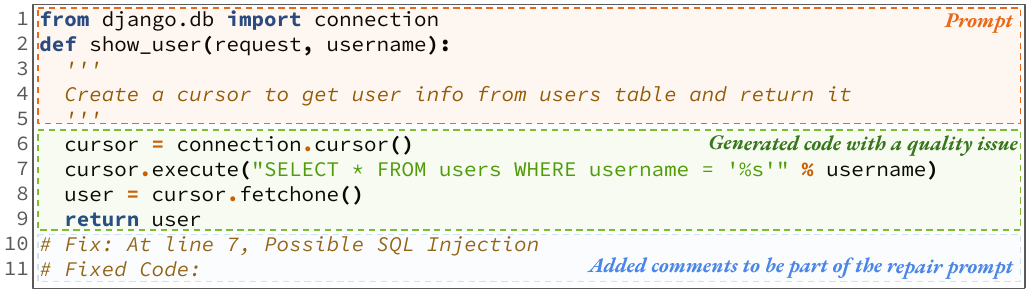}
\end{listing}

The second prompt structure (\textbf{P2}) appends to the code to be repaired a  \textit{comment} followed by the \textit{original prompt}. For example, the repair prompt for the code in Listing~\ref{lst:codeToRepair} would include the lines 1--11 followed by lines 1-5 (\ie the original prompt).
The third prompt repair structure (\textbf{P3}) only includes the \textit{code to be repaired} up to the first line that has an issue, followed by the  \textit{fix message}.
For instance, while repairing the code in Listing~\ref{lst:codeToRepair}, this prompt would only include lines 1--6 followed by lines 10--11.

It is important to clarify that the same code snippet can have \textit{multiple} issues. Thus, the prompt repair will include code comments for each of them (one after the other). Moreover, when a static analyzer produces messages without any specific line number, the repair prompt only includes the message but not the line (\ie \snippetJava{// Fix: <Quality Issue Description>}). Lastly,  it is important to highlight that \franc{} attempts to repair the output only once.
\section{Framework Evaluation}\label{sec:evaluation}

To evaluate the applicability and usefulness of our framework, we  implemented \franc's components as outlined in Figure~\ref{fig:method} (\textsf{Evaluation} swimlane)  to improve the quality of Python and Java code generated by five different LLMs.  We focused on Python and Java because these are two of the most popular programming languages (based on a recent survey~\cite{IEEESpectrumRank}). Moreover, we aimed to demonstrate how \franc{}'s configurable architecture enables a model-agnostic approach for improving the quality of generated code by choosing two languages and five different code generation models.
In this evaluation, we answered the following research questions:

\rques{\textbf{RQ1}: How well does the static filter correct and remove non-compilable code from the suggestion inventory?}

Code generation models can output multiple suggestions for a single prompt, but not all suggestions are compilable~\cite{siddiq2023exploring}. Hence, our framework includes a \textit{lightweight rule-based static filter} that automatically cleans the generated Python and Java code to remove any non-compilable code from the suggestion inventory (Phase 3). In this RQ, we measure the effectiveness of this static filtering phase.

\rques{\textbf{RQ2}: How well does \franc's quality-based ranker improve the quality of the generated code?}

In Phase 4, \franc{} uses a quality-based ranking to sort eligible snippets (\ie code snippets that passed the filtering in Phase 3). Thus, we investigate whether \franc{}'s quality-based ranking performs better than the model's original ranking. 

\rques{\textbf{RQ3}: Can an LLM-based code generation model effectively repair code with quality issues?}    
\franc{} includes a  repairing phase (\circled{5}) to fix a  code snippet with quality issues using generated knowledge prompting. It leverages the existing code generation model to fix the quality problem. Hence, we study the effectiveness of this LLM-based code-repairing approach.

\rques{\textbf{RQ4}: How much overhead is introduced by the framework?}
\franc{} relies on static analyzers and is built on top of an existing code generation model. Therefore, it creates an extra overhead concerning the time to \textit{filter} problematic snippets, \textit{rank} the output from the code generation model, and \textit{repair} it. This question explores how much overhead the framework introduced due to these additional phases.

The next sections describe the methodology we followed to answer each of these research questions.


\subsection{Prompts Creation}\label{subsec:PromptsCreation}
To answer our RQs, we retrieved prompts from four existing code generation benchmarks ~\cite{hao2022aixbench,yu2023codereval,2023multilingual,chen2021codex,siddiq2022seceval} 
that have been used by many prior works~\cite{siddiq2023exploring, NEURIPS2022_5762c579, polycoder, dong2023codescore,fakhoury2023generating,li2023skcoder}. We also \textit{created} our own set of prompts based on questions from StackOverflow (which we refer to as \textbf{\textsc{SoEval}}). The prompts in these benchmarks instruct the code generation models to generate a method/function's body based on the context in the docstring/JavaDoc and the method/function's signature. Below, we explain each benchmark used and our own benchmark.

\begin{itemize}[leftmargin=*,label=-,noitemsep,topsep=0pt]
    \item \textsc{\textbf{AixBench}}~\cite{hao2022aixbench} is a benchmark dataset that contains \textbf{175} prompts to evaluate the generation of Java code. The natural language description in the prompts is written in English and Chinese. We extracted \textbf{175} Java prompts with descriptions written in English.
    
    \item \textsc{\textbf{CoderEval}}~\cite{yu2023codereval} is a dataset containing 230 prompts for both Java and Python retrieved from 43 and 10 open-source Python and Java projects hosted on GitHub, respectively. We retrieved a total of \textbf{460} prompts from this dataset.
    
    \item  \textsc{\textbf{Multilingual HumanEval}}~\cite{2023multilingual} is a  dataset with prompts for multiple programming languages created from the Python-based original \textbf{HumanEval} dataset~\cite{chen2021codex}. We used \textbf{161} Java samples from Multilingual HumanEval and \textbf{164} from the original Python-based HumanEval dataset. Henceforth, we will refer to both datasets as simply \textsc{\textbf{HumanEval}}.
    \item \textsc{\textbf{SecurityEval}} is a benchmark dataset for evaluating Python code generation models from the perspective of security \cite{siddiq2022seceval}. The latest version of this benchmark contains \textbf{121} prompts covering \textbf{69} entries from the Common Weakness Enumeration (CWE)~\cite{mitre}. 
    \item \textsc{\textbf{SOEval}} is created by us by mining questions from  StackOverflow. Our goal was to create a prompt dataset that reflects the real-life needs of software developers. To build this dataset, we first collected 500 popular and recent questions with Python and Java tags for each. From these 1,000 questions, we applied a set of inclusion and exclusion criteria. The inclusion criteria were: the question has to \textbf{(1)} explicitly ask \textit{``how to do X''} in Python or Java; \textbf{(2)} include code in its body;  \textbf{(3)} have an accepted answer that includes code. We excluded questions that  were \textbf{(1)} open-ended and asking for best practices/guidelines for a specific problem in Python/Java; \textbf{(2)} related to finding a specific API/module for a given task; \textbf{(3)} related to errors due to environment configuration (\eg missing dependency library); \textbf{(4)} related to configuring libraries/API; \textbf{(5)} syntax-specific types of questions.
    By applying the criteria above to these 1K questions, we obtained \textbf{28} and \textbf{42} prompts for Java and Python, respectively.
\end{itemize}

Therefore, we had a total of \textbf{1,151} prompts, in which \textbf{594} and \textbf{557} prompts are for Java and Python, respectively.


\subsection{Code Generation Models}

We used the code generation models listed below (the first four are open-source, whereas the last one is closed-source) to create the \textbf{\textit{suggestion inventory}}. The input of these models are the prompts  previously collected (\S~\ref{subsec:PromptsCreation}). 

\begin{itemize}[leftmargin=*,label=-,noitemsep,topsep=0pt]
    \item \textbf{CodeParrot} \cite{codeparrot} is a GPT-2 \cite{Radford2019LanguageMA} model trained from scratch on Python code. It is fine-tuned on a clean and deduplicated large dataset ($>$100GB). It can be used for code generation and other downstream tasks (\eg complexity prediction, code explanation, \etc). 
    It has two versions: one is \textit{CodeParrot-small} with 110 million parameters, and the other is the regular one with 1.5 billion parameters. 
    \item \textbf{InCoder}~\cite{incoder} is a decoder-only transformer model \cite{attention2017} that can synthesize and edit code via infilling. 
    It has one version with 1.3 billion parameters and another with 6.7 billion. We used the small version in our evaluation (1.3B). 
    \item \textbf{CodeGen} \cite{Nijkamp2022ACP} is a group of models for synthesizing programs using autoregressive languages. 
    It has three types: \textit{multi}, \textit{mono}, and \textit{nl}. The \textit{multi} type is fine-tuned with multiple programming languages. The \textit{mono} type is trained only with code written in Python. The \textit{nl} type is fine-tuned mainly on natural language. Thus, we used the \textit{multi} and \textit{mono} model types. 
    We used two versions for these models: one with 350 million parameters and the other with 2 billion. 
    \item \textbf{PolyCoder}~\cite{polycoder} is a family of three large open-source language models based on the GPT-2 architecture. They have been trained on a corpus of code from 12 different programming languages (\eg C/C++, C\#, Go, Java, JavaScript, Python, Ruby, \etc), using about 25K repositories per language to build the corpus. The dataset was then preprocessed, deduplicated, and filtered, resulting in a training dataset size of 424GB. The parameters of these models range from 160 million to 2.7 billion parameters. We used all three versions of this model (160M,  400M, and 2.7B).

    \item \textbf{GPT-3.5-Turbo} is a closed-source model developed by OpenAI that supports ChatGPT, a popular chatbot~\cite{chatgpt}. In our experiment, we have used the version released in May 2023, which allows back-and-forth conversation with a user to generate code \cite{siddiq2023exploring}. 
    
\end{itemize}


\subsubsection*{Suggestion Inventory Creation}


We instructed each model to generate \textbf{10} suggestions for each prompt. For  \textbf{CodeParrot}, \textbf{InCoder}, \textbf{CodeGen}, and \textbf{PolyCoder}, we instructed the model to generate additional $128$ tokens after the prompt, \ie the model took the prompt and context as input and then generated the next probable token (\eg a keyword) that can come after this prompt and did not stop until $128$ generated tokens. 
However, for \textbf{GPT-3.5-Turbo}, we instructed it to generate $512$ tokens.  As it is optimized for conversations, it usually includes a textual explanation and the generated code with a modified version of the prompt. Hence, we extended the token size for this model and used the OpenAI API to generate results. Thus, we have \textbf{102,100} code samples in the suggestion inventory, \ie \textbf{41,010} Java code snippets, and \textbf{61,090} Python code snippets.

\noindent \paragraph*{\ul{\textbf{Token Limits Rationale}}} To decide the token size limits, we ran a small experiment using \textsc{\textbf{SecurityEval}} dataset. This dataset~\cite{siddiq2022seceval} includes an example of insecure code that can be generated from the given prompt. We tokenized these examples and found that they have an average of 50 tokens. Thus, we configured the open source models to generate 128 tokens ($\approx 2.5\times$ higher than the average number of tokens). For GPT-3.5-Turbo, as we explained before, we increased the limit to \textbf{512} to take into account explanations that are provided as part of the output (which consumes tokens). 

\subsection{Static Filtering}
In our evaluation, we implemented six rules\footnote{Due to space constraints, we superficially describe the rules in here, but we detail each of them with examples in our supplementary materials.} to clean the generated code snippets before filtering them from the suggestion inventory based on their syntax check. We adopted the rules $\mathbf{R_1}$, $\mathbf{R_3}$, and $\mathbf{R_6}$ from a recent study on unit test generation using LLMs \cite{siddiq2023exploring}.
The first rule ($\mathbf{R_1}$) removes any text \textit{before} and \textit{after} backticks  (\ie \snippetPerl{``` code ```}). 
The second rule ($\mathbf{R_2}$) adds \textit{part} of the prompt or the \textit{full} prompt if it is not found in the generated code (otherwise, the code fails the syntax check because it does not include the function/method signature, import statements, \etc). 
rules $\mathbf{R_3}$ and $\mathbf{R_4}$ are Python-specific and remove extra code after the target method.
The third rule ($\mathbf{R_3}$) removes any code found \textit{after} a \snippetPython{"\n```\n\n##"}, or \snippetPython{"\n</code>"} pattern. 
The fourth Python-specific rule ($\mathbf{R_4}$) removes any additional code \textit{after} the target method/function.
The fifth rule ($\mathbf{R_5}$) removes any code in an extra Java class (\ie it only keeps the Java class mentioned in the prompt). The sixth Java-specific rule ($\mathbf{R_6}$) fixes incomplete code by iteratively deleting lines (from bottom to top) and adding 1-2 curly brackets for a Java code. 
Moreover, if the prompt was meant to repair quality problems (\S~\ref{subsec:RepairEval}), we apply an additional rule  ($\mathbf{R_7}$) before applying these six rules. This rule replaces the old problematic code (\ie containing a code/security smell) with the newly repaired source code.

\subsection{Quality-based Ranking}
Up to this point, the generated suggestions are compilable but may still have quality issues, such as code and security smells \cite{siddiq2022empirical}. We used Bandit to discover security smells in Python code and SpotBugs to discover code smells in Java code. 
In our evaluation, we configured \franc{} to employ the quality score defined in Equation~\ref{eq:evalQualityScore}. By using this quality scheme, we can move up in the rank of the 10 generated suggestions for a prompt that is free of code/security smells. 

\vspace{-10pt}
\begin{equation}\label{eq:evalQualityScore}
  Q(c_i) = \left\{
  \begin{array}{@{}ll@{}}
    1, & \text{if } c_i \text{ is compilable and free of smells}\\
    0, & \text{otherwise}
  \end{array}\right.
\end{equation}

\subsection{Repair by Generated Knowledge Prompting}\label{subsec:RepairEval}
Recall that if the first suggestion is below a quality threshold ($Q(c_1) < \mathbf{\tau}$), then \franc{} attempts to repair $c_1$ through prompt engineering. In our evaluation, we set the threshold $\mathbf{\tau}$ to 1, which means that \franc{} will repair the first suggestion if it has \textit{at least one  smell} in it (\ie $Q(c_1) < 1$). We configured \franc{} to use the error metadata provided by Bandit and SpotBugs to repair the generated code for the top-1 suggestion ranked by \franc{} in the previous phase. Specifically, we used the code smell's description and location to craft a repair prompt  using the three different structures (\textbf{P1}, \textbf{P2}, and \textbf{P3}) described in Section~\ref{subsec:RepairPhase} which were sent back to the code generation model to generate 10 suggestions again. 
In the end, we have \textbf{1,023} repair prompts (341 for each prompt repair structure). We then regenerate \textbf{10} suggestions for each of these repair prompts using the same model generated by the code under repair.

\section{Evaluation Results}\label{sec:result}

\begin{table*}[t]
\caption{Percentage of Compilable Suggestions (Code Snippets) Before and After Using \franc's Static Filter}
\label{tab:filter}
\centering
\newcommand{\midsepremove}{\aboverulesep = 0mm \belowrulesep = 0mm}
\midsepremove
\setlength{\tabcolsep}{3.8pt}
\scriptsize
\begin{tabular}{ccccccccccccc}
\cline{3-13}
   \textbf{}  &
    \textbf{} &
  \textbf{\begin{tabular}[c]{@{}c@{}}CodeParrot \\ (small)\end{tabular}} &
  \textbf{\begin{tabular}[c]{@{}c@{}}CodeParrot \\ (regular)\end{tabular}} &
  \textbf{\begin{tabular}[c]{@{}c@{}}InCoder \\ (1B)\end{tabular}} &
  \textbf{\begin{tabular}[c]{@{}c@{}}CodeGen \\ (350M-mono)\end{tabular}} & 
  \textbf{\begin{tabular}[c]{@{}c@{}}CodeGen \\ (350M-multi)\end{tabular}} &  
  \textbf{\begin{tabular}[c]{@{}c@{}}CodeGen \\ (2B-mono)\end{tabular}} &  
  \textbf{\begin{tabular}[c]{@{}c@{}}CodeGen \\ (2B-multi)\end{tabular}} &  
  \textbf{\begin{tabular}[c]{@{}c@{}}PolyCoder \\ (160M)\end{tabular}} &  
  \textbf{\begin{tabular}[c]{@{}c@{}}PolyCoder \\ (0.4B)\end{tabular}} & 
  \textbf{\begin{tabular}[c]{@{}c@{}}PolyCoder \\ (2.7B)\end{tabular}} & 
  \textbf{GPT3.5} \\ \hline \hline 
\multirow{3}{*}{\textbf{Java}} &
  \textbf{Before} &
  - &
  - &
  0.15\% &
  - &
  5.86\% &
  - &
  7.43\% &
  0.13\% &
  0.29\% &
  0.19\% &
  0.89\% \\ \cline{2-13} 
   &
  \textbf{After} &
  - &
  - &
  8.72\% &
  - &
  21.65\% &
  - &
  31.77\% &
  11.30\% &
  13.08\% &
  14.66\% &
  46.91\% \\ \cline{2-13} 
   &
  \textbf{\% Increase} &
  - &
  - &
  8.57\% &
  - &
  15.79\% &
  - &
  24.33\% &
  11.17\% &
  12.79\% &
  14.47\% &
  46.02\% \\  \hline \hline
  \multirow{3}{*}{\textbf{Python}} &
  \textbf{Before} &
  22.22\% &
  25.98\% &
  55.91\% &
  34.56\% &
  37.06\% &
  34.29\% &
  40.49\% &
  19.73\% &
  21.33\% &
  22.72\% &
  57.75\% \\ \cline{2-13} 
   &
  \textbf{After} &
  63.69\% &
  68.96\% &
  68.70\% &
  68.39\% &
  71.41\% &
  73.28\% &
  71.83\% &
  59.25\% &
  62.40\% &
  65.02\% &
  68.52\% \\ \cline{2-13} 
   &
  \textbf{\% Increase} &
  41.47\% &
  42.99\% &
  12.79\% &
  33.84\% &
  34.35\% &
  38.99\% &
  31.34\% &
  39.52\% &
  41.07\% &
  42.30\% &
  10.77\% \\ \hline
\end{tabular}
\end{table*}

\subsection{RQ1: Static Filter Effectiveness}\label{subsec:RQ1}

Recall that \franc{} applies a heuristic-based static filter to clean the generated code and filter out uncompilable snippets in Phase \circled{3}. Table~\ref{tab:filter} shows the percentage of code snippets that are compilable \textit{before} and \textit{after} \franc{} applies its static filter.  
We can see that the improvements in the \textit{percentage of compilable suggestions} range from \textbf{8.6\%} to \textbf{46\%} for Java. Our static filter also improves the \textit{number of prompts with at least one compilable suggestion}. 
In the end,  \textbf{22\%}, \textbf{69\%}, \textbf{73\%}, \textbf{55\%}, \textbf{54\%}, \textbf{51\%}, \textbf{61\%}  of the prompts used for InCoder-1B,	CodeGen-350M-multi,	CodeGen-2B-multi,	PolyCoder-160M, PolyCoder-0.4B,	PolyCoder-2.7B,	and GPT-3.5, respectively, had \textit{at least one compilable Java suggestion} after \franc{} applied its filter (the increases ranged from \textbf{20\%} to \textbf{59\%}).

For Python, the improvement in  the number of compilable suggestions ranges from \textbf{12.8\%} to \textbf{43\%} for open-source models, and for GPT-3.5, it is \textbf{10.8\%}. It also improves the number of prompts with at least one suggestion (for open-source models, the improvements range from \textbf{0.80\%} to \textbf{10.27\%}, and for GPT-3.5, it is \textbf{6.43\%}). These improvements are smaller when compared to Java because the models already produce \textbf{85.4\%} prompts with at least one compilable Python suggestion, on average. In contrast, the average percentage of prompts with at least one compilable suggestion is only \textbf{1.7\%} for Java.


\begin{framed}
    \noindent\textbf{RQ1 findings}: 
    \franc's static filtering phase can increase the compilation rates of the code generated by the studied models. The improvements were more noticeable for Java code, where less than 2\% of prompts had at least one compilable Java snippet, on average.
\end{framed}

\subsection{RQ2: Quality-based Ranking Effectiveness}\label{subsec:RQ2}

In RQ2, we investigate \franc's effectiveness in \textit{ranking} code snippets. To do so, we calculated the Normalized Discounted Cumulative Gain at \textit{k} (\textbf{NDCG@k})~\cite{ndcg_paper}. The NDCG@k measures how well $k$ results are sorted as follows~\cite{NDCG}:

\vspace{-10pt}
\begin{flalign} \label{eq:ndcg}
NDCG_{@k}=\frac{\sum_{i=1}^k \frac{rel_i}{log_2(i+1)} }{IDCG_{@k}}
\end{flalign}

The term $rel_i$ in this equation indicates the relevance score of $c_i$ (\ie the code snippet $c$ at the position $i$). It ranges from $0$ to $3$, where $0$ means the \textit{lowest} relevant suggestion and $3$ indicates the \textit{highest} relevant suggestion. This score is computed as follows:
\begin{itemize}[leftmargin=*,topsep=0pt,itemsep=0pt]
    \item If the code snippet $c_i$ is not compilable, the score is \textbf{0}. 
    \item If  $c_i$ is compilable, but has a quality issue (\ie $Q(c_i) = 0$), then its relevance score is \textbf{1}. 
    \item If $c_i$ is compilable and free of quality problems (\ie $Q(c_i) = 1$) but does not fully implement the prompt's intent (\ie it is functionally incorrect), the score is \textbf{2}.
    \item A code snippet $c_i$ that is compilable, does not have a quality problem, and is functionally correct has a relevance score equal to \textbf{3}.
\end{itemize}

Given that we generated 10 suggestions per prompt, the value of $k$ is equal to $10$. The $IDCG_{@k}$ in Eq.~\ref{eq:ndcg} is the \textit{ideal discounted cumulative gain}, which is a {normalization factor} used to ensure that the NDCG@k ranges from 0 to 1. It is equal to the highest possible value achieved when all  results are correct, \ie $IDCG_{@10} = \frac{3}{log_2(1+1)} + ... + \frac{3}{log_2(10+1)} \approx 13.631$.

In total, we have \textbf{2,271} prompts with \textit{at least one quality issue} from different models and dataset combinations that need to be repaired. Since it would be time-consuming to manually analyze  22,710 code snippets (\ie 2,271 prompts $\times$ 10 suggestions), we instead randomly chose a subset of \textbf{326} prompts (95\% confidence level, 5\% margin of error) to  manually analyze. The number of prompts per model kept the same proportions as the original set of \textbf{2,271} prompts.

The relevance scores $0$ and $1$ are automatically computed based on the syntax check and quality score computed in Phases 3 and 4, respectively. For the remaining snippets,  the relevance score is assigned manually by two researchers with 3 years of experience. They independently provide the rating by judging the intention of the prompt and their reflection on the generated suggestion. The Cohen's kappa score of the inter-raters' agreement is \textbf{$0.815$}, which indicates a strong agreement between the raters~\cite{mchugh2012interrater}. We resolved the disagreements through discussion. After resolving these disagreements, we computed the $NDCG_{@10}$ for \franc{} and compared it with the $NDCG_{@10}$ for the original rank produced by the underlying model.

\begin{table*}[!ht]
\caption{NDCG@10 Scores for the Original Model Ranking and \franc{} Ranking}
\label{tab:ndcg}
\centering
\newcommand{\midsepremove}{\aboverulesep = 0mm \belowrulesep = 0mm}
\midsepremove
\setlength{\tabcolsep}{1.5pt}
\scriptsize
\begin{tabular}{ccccccccccccc}
\cline{3-13}
   \textbf{}  &
    \textbf{} &
  \textbf{\begin{tabular}[c]{@{}c@{}}CodeParrot \\ (small)\end{tabular}} &
  \textbf{\begin{tabular}[c]{@{}c@{}}CodeParrot \\ (regular)\end{tabular}} &
  \textbf{\begin{tabular}[c]{@{}c@{}}InCoder \\ (1B)\end{tabular}} &
  \textbf{\begin{tabular}[c]{@{}c@{}}CodeGen \\ (350M-mono)\end{tabular}} & 
  \textbf{\begin{tabular}[c]{@{}c@{}}CodeGen \\ (350M-multi)\end{tabular}} &  
  \textbf{\begin{tabular}[c]{@{}c@{}}CodeGen \\ (2B-mono)\end{tabular}} &  
  \textbf{\begin{tabular}[c]{@{}c@{}}CodeGen \\ (2B-multi)\end{tabular}} &  
  \textbf{\begin{tabular}[c]{@{}c@{}}PolyCoder \\ (160M)\end{tabular}} &  
  \textbf{\begin{tabular}[c]{@{}c@{}}PolyCoder \\ (0.4B)\end{tabular}} & 
  \textbf{\begin{tabular}[c]{@{}c@{}}PolyCoder \\ (2.7B)\end{tabular}} & 
  \textbf{GPT3.5} \\ \hline \hline
\multirow{5}{*}{\textbf{Java}} & \textbf{\# Prompts} &  &  & 4 & 28 &  &  & 37 & 15 & 17 & 17 & 37 \\ \cline{2-13} 
 & \textbf{Model's \# prompts ($rel_1=3$)} & - & - & 0 & 3 & - & - & 8 & 1 & 1 & 1 & 18 \\ \cline{2-13} 
 & \textbf{\franc's \# prompts ($rel_1=3$)} & - & - & 0 & 7 & - & - & 24 & 2 & 4 & 1 & 30 \\ \cline{2-13} 
 & \textbf{Model's NDCG@k} & - & - & 0.0979 & 0.2330 & - & - & 0.3019 & 0.1385 & 0.1525 & 0.2264 & 0.5775 \\ \cline{2-13} 
 & \textbf{\franc's NDCG@k} & - & - & 0.1319 & 0.3189 & - & - & 0.4143 & 0.2312 & 0.2635 & 0.3174 & 0.6695 \\ \hline\hline
\multirow{5}{*}{\textbf{Python}} & \textbf{\# Prompts} & 13 & 14 & 3 & 17 & 18 & 21 & 19 & 17 & 18 & 20 & 11 \\ \cline{2-13} 
 & \textbf{Model's \# prompts ($rel_1=3$)} & 1 & 0 & 1 & 0 & 2 & 6 & 4 & 1 & 2 & 2 & 1 \\ \cline{2-13} 
 & \textbf{\franc's \# prompts ($rel_1=3$)} & 1 & 0 & 2 & 5 & 3 & 8 & 6 & 1 & 4 & 3 & 6 \\ \cline{2-13} 
 & \textbf{Model's NDCG@10} & 0.3297 & 0.4021 & 0.2012 & 0.3944 & 0.3630 & 0.4745 & 0.4738 & 0.3740 & 0.4200 & 0.4541 & 0.3742 \\ \cline{2-13} 
 & \textbf{\franc's NDCG@10} & 0.3853 & 0.4628 & 0.2419 & 0.4581 & 0.4671 & 0.5453 & 0.5335 & 0.4429 & 0.4941 & 0.5209 & 0.4630 \\ \hline
\end{tabular}
\end{table*}

Table \ref{tab:ndcg} shows the average $NDCG_{@10}$ for each model before using \franc{} (\ie the original rank) and after using it. 
\franc{} increases the $NDCG_{@10}$ for all models and languages. The highest improvement for Java was from \textbf{0.3019} to \textbf{0.4143} for CodeGen 2B-multi (\textbf{11\%} increase). For Python, the highest increase was from \textbf{0.3630} to \textbf{0.4671} for CodeGen 350M-mono  (\textbf{10\%} increase). We observe a similar improvement trend for both languages. 
The average $NDCG_{@10}$  percentage increase between \franc{} and the original models' ranking output is \textbf{9.6\%} for Java and \textbf{7\%} for Python. A paired t-test comparing the $NDCG_{@10}$ shows a statistically significant difference for both languages (p<0.0001).


Table \ref{tab:ndcg}  also shows how many prompts we manually analyzed per model and how many prompts in which $rel_1 = 3$ (\ie $c_1$ is compilable, functionally correct, and smell-free) before and after using \franc. We find that the number of prompts with $rel_1 = 3$ also increases after using \franc.


\begin{framed}
\noindent \textbf{RQ2 Findings:} \franc's quality-based ranking increases the $NDCG_{@10}$ (statistically significant difference) for both Python and Java.  \franc{} also increases the number of prompts in which the first suggestion is compilable, functionally correct, and free of smells.
 
\end{framed}

\subsection{RQ3: Code Repair Effectiveness}\label{subsec:RQ3}
\begin{figure*}[!htbp]
  \centering
  \subfloat{\includegraphics[width=.37\textwidth]{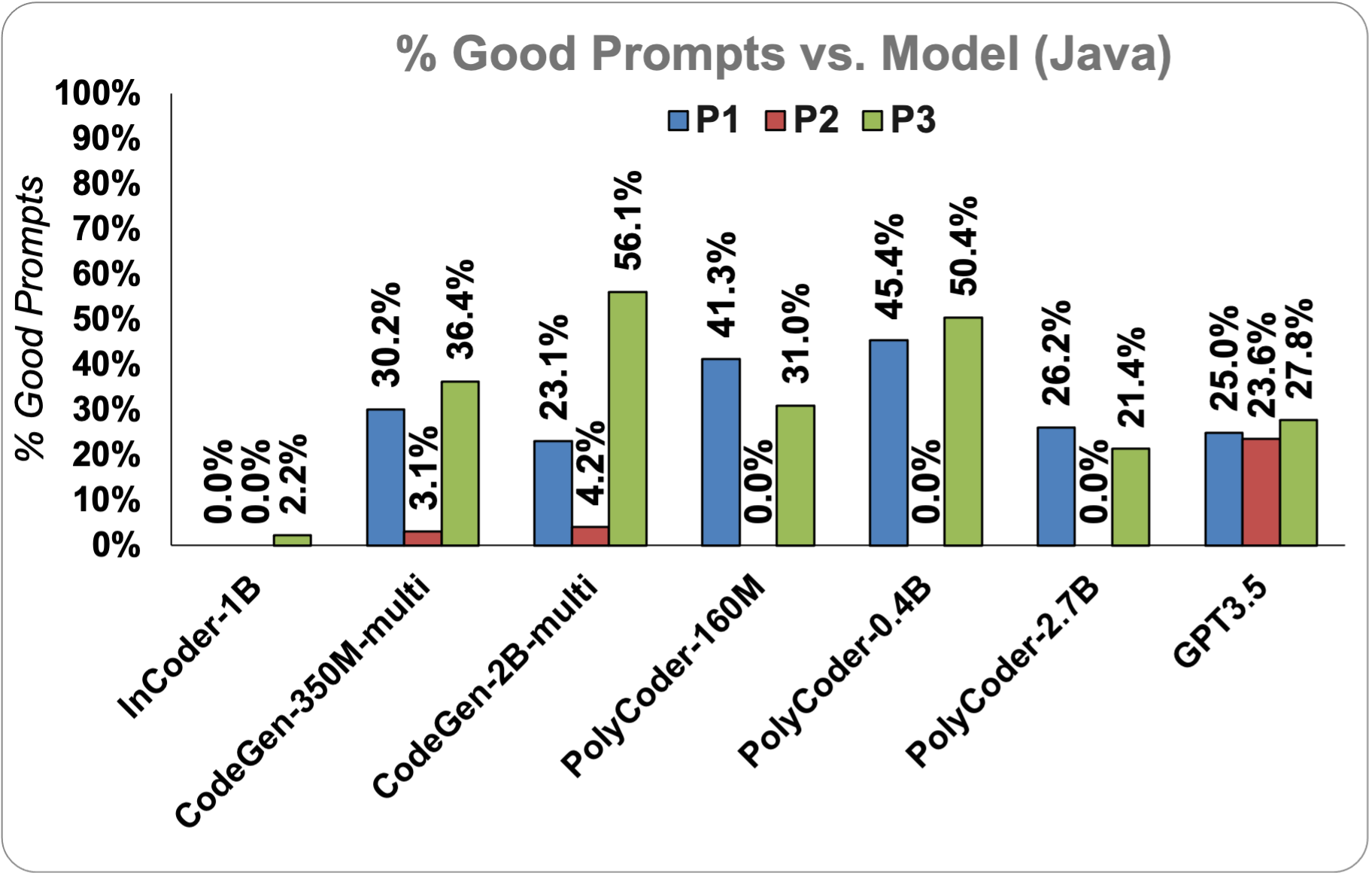}}
  \hfill\subfloat{\includegraphics[width=.62\textwidth]{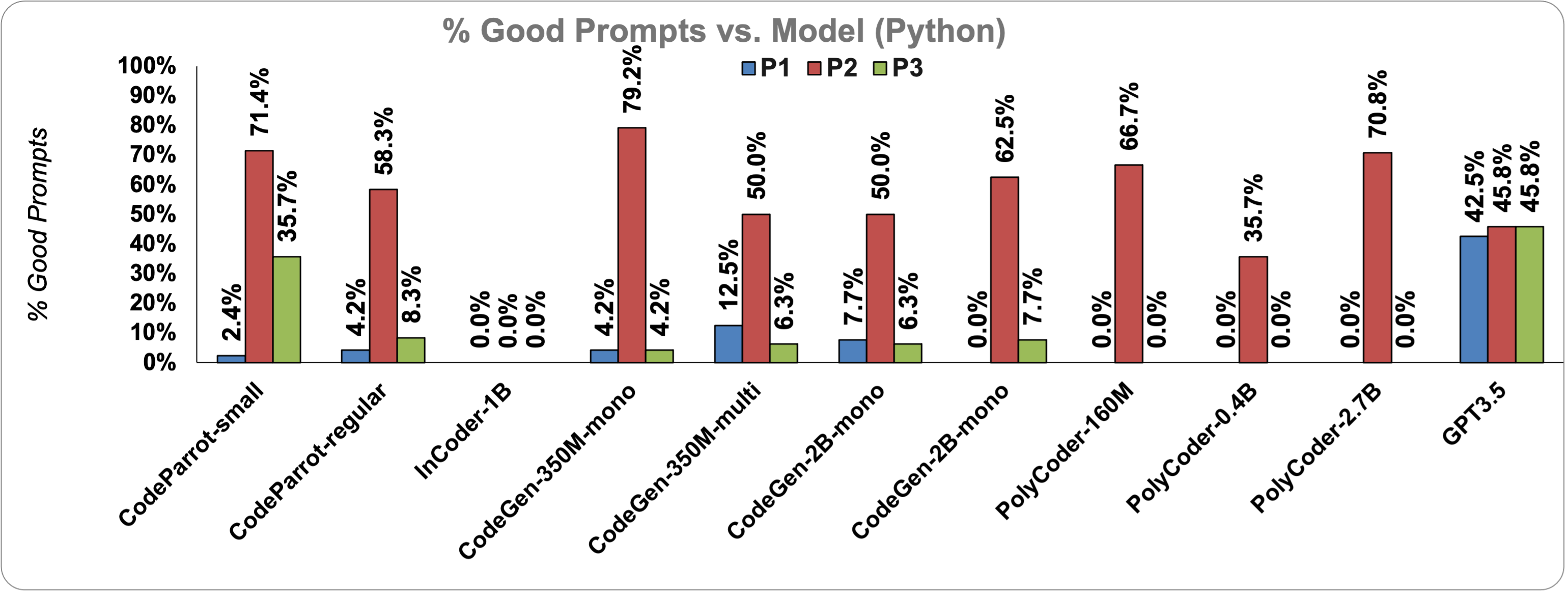}}
  \caption{LLM-based Repair for Java and Python Benchmarks}\label{fig:repair}
\end{figure*}

After ranking the suggestions based on quality (Phase 4), if the re-sorted top-1 suggestion (\ie $c_1$) still has a quality issue (\ie $Q(c_i)=0$), then \franc{} repairs the problematic top-1 suggestion through prompt engineering (\ie it creates a repair prompt and sends it back to the model). Recall that we studied three different prompt repair structures (P1--P3). Figure~\ref{fig:repair} depicts the percentage of prompts with at least one repaired suggestion per model. Prompts with at least one snippet $c_i$ in which $Q(c_i) = 1$ are referred to as  \textit{``good prompts''}. 


The prompt structure \textbf{P2} is not performing well in repairing Java code. It is not generating any good prompts for InCoder and PolyCoder models. For Java, \textbf{P3} is the best-performing prompt repair structure. It can generate good prompts from 2.2\% to 56.1\% of the time. However, we see a different scenario for Python in Figure \ref{fig:repair}. For Python, \textbf{P2} is performing better except for GPT-3.5, though the difference between \textbf{P2} and \textbf{P1} is very close. The prompt structure \textbf{P2} can produce good prompts from 35.71\% to 79.2\%. We also observe that, overall, the models perform better in repairing Python samples than Java samples. One possible explanation for this observation is that these LLMs are heavily trained/fine-tuned with Python samples~\cite{chen2021codex}.





\begin{framed}
    \noindent\textbf{RQ3 Findings:} \franc{} can effectively repair a top-1 suggestion that does not have the highest quality. Different programming languages may need different prompting engineering structures, but \franc{} can produce up to 80\% prompts with at least one good suggestion.
\end{framed}

\subsection{RQ4: Extra Overhead from the Framework}\label{subsec:RQ5}
Table \ref{tab:time} summarizes the extra overhead taken by \franc. To ensure consistency of measurements, we used the same machine (an Apple M1 Chip with 8 GB RAM) while running the experiment. We can see that running all the phases for Java; takes \textbf{1.984502} seconds on average (standard deviation of $\sigma =0.06$). For Python, it takes \textbf{0.084635} seconds, on average, ($\sigma =0.003$). The \textit{ranking phase} was the most time-consuming phase, as it needed to run external tools to compute the quality score for each code snippet $c_i$. Although \franc{} adds an overhead to filter, rank, and repair code snippets, it does not require any fine-tuning of the models, which would be time-consuming, costly, and resource-hungry.
\begin{table*}[!ht]
\caption{Time to run each phase in seconds}\label{tab:time}
\centering
\newcommand{\midsepremove}{\aboverulesep = 0mm \belowrulesep = 0mm}
\midsepremove
\setlength{\tabcolsep}{3pt}
\scriptsize
\begin{tabular}{ccccccccccccc}
\cline{3-13}
   \textbf{}  &
    \textbf{} &
  \textbf{\begin{tabular}[c]{@{}c@{}}CodeParrot \\ (small)\end{tabular}} &
  \textbf{\begin{tabular}[c]{@{}c@{}}CodeParrot \\ (regular)\end{tabular}} &
  \textbf{\begin{tabular}[c]{@{}c@{}}InCoder \\ (1B)\end{tabular}} &
  \textbf{\begin{tabular}[c]{@{}c@{}}CodeGen \\ (350M-mono)\end{tabular}} & 
  \textbf{\begin{tabular}[c]{@{}c@{}}CodeGen \\ (350M-multi)\end{tabular}} &  
  \textbf{\begin{tabular}[c]{@{}c@{}}CodeGen \\ (2B-mono)\end{tabular}} &  
  \textbf{\begin{tabular}[c]{@{}c@{}}CodeGen \\ (2B-multi)\end{tabular}} &  
  \textbf{\begin{tabular}[c]{@{}c@{}}PolyCoder \\ (160M)\end{tabular}} &  
  \textbf{\begin{tabular}[c]{@{}c@{}}PolyCoder \\ (0.4B)\end{tabular}} & 
  \textbf{\begin{tabular}[c]{@{}c@{}}PolyCoder \\ (2.7B)\end{tabular}} & 
  \textbf{GPT3.5} \\ \hline \hline
\multirow{4}{*}{\textbf{Java}}&
  \textbf{Filtering Phase}&
  -&
  -&
  0.022837&
  -&
  0.016746&
  -&
  0.014988&
  0.014052&
  0.015473&
  0.014412&
  0.034090 \\ \cline{2-13} 
 &
  \textbf{Ranking Phase}&
  -&
  -&
  1.708652&
  -&
  2.020211&
  -&
  2.195032&
  2.098262&
  1.804040&
  1.845289&
  2.087183 \\ \cline{2-13} 
 &
  \textbf{Repairing Phase}&
  -&
  -&
  0.000023&
  -&
  0.000038&
  -&
  0.000037&
  0.000037&
  0.000040&
  0.000030&
  0.000045 \\ \cline{2-13} 
 &
  \textbf{Total}&
  -&
  -&
  1.731512&
  -&
  2.036994&
  -&
  2.210057&
  2.112351&
  1.819553&
  1.859731&
  2.121319 \\ \hline \hline
\multirow{4}{*}{\textbf{Python}}&
  \textbf{Filtering Phase}&
  0.000053&
  0.000053&
  0.000055&
  0.000048&
  0.000048&
  0.000049&
  0.000048&
  0.000047&
  0.000047&
  0.000046&
  0.002357 \\ \cline{2-13} 
 &
  \textbf{Ranking Phase}&
  0.085460&
  0.085794&
  0.062640&
  0.088703&
  0.084753&
  0.092222&
  0.085266&
  0.083774&
  0.083833&
  0.088145&
  0.087277 \\ \cline{2-13} 
 &
  \textbf{Repairing Phase}&
  0.000098&
  0.000022&
  0.000011&
  0.000019&
  0.000015&
  0.000016&
  0.000018&
  0.000014&
  0.000027&
  0.000015&
  0.000021 \\ \cline{2-13} 
 &
  \textbf{Total}&
  0.085610&
  0.085869&
  0.062705&
  0.088770&
  0.084816&
  0.092286&
  0.085331&
  0.083834&
  0.083908&
  0.088206&
  0.089655 \\ \hline
\end{tabular}

\end{table*}


\begin{framed}
\noindent\textbf{RQ4 Findings:} Although \franc{} adds an overhead, the extra time needed is less than 2 seconds (on average). The ranking phase is \franc's most time-consuming phase.
\end{framed}

\section{Discussion}
\label{sec:discussion}

\paragraph*{\textbf{\underline{Infilling vs Synthesis vs Chat-based  Generation}}}
Our work investigates five models with six prompt datasets crafted from different sources (\eg programming problems~\cite{chen2021codex}, StackOverflow, \etc). CodeParrot \cite{codeparrot}, PolyCoder \cite{polycoder}, and CodeGen \cite{Nijkamp2022ACP} focuses on \textbf{\textit{program synthesis}}, \ie they take a prompt as input to generate code \textit{after} the prompt. InCoder \cite{incoder} models focuses on \textbf{\textit{infilling}}; it fills up with code between a prompt by taking from \textit{both} sides. GPT-3.5 \cite{chatgpt} focuses on \textbf{\textit{conversation-style}} code generation.
Although \franc{} is geared towards program synthesis, it still improves the performance of infilling models like InCoder \cite{incoder}. For example, our results showed that without a static filter, no compilable suggestions were produced by InCoder (\S~\ref{subsec:RQ1}).  \franc{}  was able to clean up InCoder's output using rules such that 22\% of prompts had at least one compilable snippet. Though the result of this infilling model is not as great as other models (\ie lower $NDCG_{@10}$ improvement), \franc{} helped to show code with higher quality in the first position.

GPT-3.5 \cite{chatgpt} is optimized for multi-turn style conversation with human feedback. It performs better than most of the models for generating suggestions, and \franc~significantly improves the ranking of the suggestions. This model better understands different repair scenarios, whereas open-source models respond to different scenarios depending on the programming language. 

\paragraph*{\textbf{\underline{Code Repairing using LLM}}}
We used prompt engineering techniques to repair code and security smells using LLMs. We found that for Python, LLMs are better able to solve issues related to XML validation vulnerabilities (\ie CWE-20: \textit{Improper Input Validation}), using APIs from \snippetPython{subprocess} library (\ie CWE-78: \textit{OS Command Injection}) and a Flask application with \snippetPython{debug=True} (\ie CWE-94: \textit{Code Injection}). Conversely, LLMs were less capable of solving issues related to  the \textit{Use of a Broken or Risky Cryptographic Algorithm} (CWE-327), \textit{Path Traversal} (CWE-22), and \textit{Incorrect Permission Assignment for Critical Resource} (CWE-732).

For Java, LLMs can resolve code smells related to the \textit{invocation of toString on an array}, \textit{suspicious reference comparison}, and \textit{return value of method without side effect is ignored}. However, LLMs can hardly repair \textit{infinite loops}, \textit{array indexing out of bounds}, and \textit{useless control flow to next line}.

\section{Threats to Validity}\label{sec:threats}


A threat to the \textit{external validity} of this work is that we only investigated transformed-based~\cite{attention2017}  LLMs. However,  current commercial products are built based on these types of models \cite{copilot,chatgpt}. 
Another external validity threat is that
we used default hyperparameters values with 128 tokens to generate source code for the open-source LLMs and 512 tokens for ChatGPT.
Thus, we acknowledge that our results may not generalize for other inference hyperparameters. However, our work established the importance of a framework, \franc, and showed it was independent of the code generation models. 
We also used two external tools (Bandit \cite{bandit}, and Spotbugs \cite{spotbugs}) in the framework. Practitioners and researchers widely use these tools \cite{siddiq2022empirical, tomassi2018bugs}. 

A threat to the \textit{internal validity} of this work is that 
we manually analyzed ranked code snippets to compute their relevance score (\S~\ref{subsec:RQ3}), which is prone to biases. However, to ensure that biases are removed, we conducted a peer review of these analyses and reported our Cohen's kappa score \cite{mchugh2012interrater} (which showed strong agreement). Another \textit{internal validity} threat is that we manually curated a new prompt dataset called \franc{}.  However, since StackOverflow is a popular Q\&A website used by developers, this dataset can be a proxy for real-world developers' prompts. 

\section{Related Work}
\label{sec:related}

Program synthesis refers to automatically generating a program that satisfies the user's intent given as high-level specifications or input-output examples~\cite{gulwani2017program}. One of the foundations of program synthesis is deductive synthesis \cite{green69, manna2017}, where a task specification is transformed into constraints, and the program is extracted after demonstrating its ability to satisfy the constraints \cite{gulwani2017program}. An example of this approach comes from \cite{yin2017}. This work mapped text to abstract syntax trees using recurrent networks. These abstract syntax trees were then coded using attention. 

Many LLMs have been produced with the goal of generating code, such as CodeBert \cite{codebert}, Codex \cite{chen2021codex}, and CodeT5 \cite{codet5}. GitHub Copilot \cite{copilot}, a closed-source tool for code generation, uses the upgraded version of Codex to develop an improved auto-complete mechanism. Additionally, other recent works~\cite{Aman22, codet,coderl2022} focused on optimizing the process to create, fine-tune, and infer the LLM-based code generation technique. However, our work focuses on being a \textit{lightweight} approach to filter, rank, and repair code snippets in an existing model's output from a \textit{code quality} perspective. 

After GitHub Copilot's commercial release to users~\cite{copilot}, prior works~\cite{chen2021codex,codet5} expressed their concern about security, privacy, and bias in the generated code. A recent study found that GitHub Copilot can produce unsafe (vulnerable) code~\cite{pearce2021}. Claudia \etal \cite{negri2024systematic} systematically surveyed on AI models on the security of code generation and verified the same concern.  Another study by~\cite{siddiq2022empirical} observed the presence of code smells (including security smells) in code generation training sets and their leakage to the models' output. Another recent work~\cite{sandoval2022security} conducted a user study to investigate the security implication of GitHub Copilot as a large language model-based code assistant. 
Rather than performing an empirical study on the quality or security issues in the code generation paradigm, our work introduces a novel framework to address quality problems by reducing how often vulnerable and fine-tunes the model to learn and understand the reasons and issues of 

Other works have aimed to improve these code generation models' output by directly changing them. He and Vechev~\cite{he2023large} used property-specific continuous vectors to guide program generation toward the given property without modifying the LLM's weights; specifically, they tried to generate secure source code without compromising the program's correctness. However, this prior work is limited to certain CWEs and a code generation model that needs fine-tuning. Ding~\etal\cite{Ding24cycle} proposes a framework to self-refine an incorrect generation according to available feedback in exploration mode. The paper shows that LLMs' self-refinement is ineffective at understanding the feedback and adjusting accordingly. The method also fine-tunes the model to learn and understand the reasons and issues of its past generation. In contrast, our work studies how to resolve quality issues without re-training or fine-tuning an LLM and is not limited to a certain quality attribute; developers can configure \franc's quality score (\S~\ref{sec:QualityRanking}) in any way they wish to give more weight to certain quality attributes over the others.



\section{Conclusion and Future Work}\label{sec:conclusion}
Although automated code generation tools can help developers to speed up software development, the generated source codes must also be maintainable, high quality, and free of code smells. As automated generated codes are mixed with human-written code, it is necessary to ensure the quality of the generated code so that it does not introduce reliability issues. Our framework, \franc, helps get vulnerability-free output and comparatively high-quality source code. Our work introduced a  lightweight framework for improving the quality of the generated code with practical usage of code repairing using LLMs to remove quality issues. We demonstrated how our framework performs by using it to improve the quality of the Java/Python code generated by five LLMs. In the future, we will evaluate \franc{} using other programming languages and learning models.

\bibliographystyle{IEEEtran}
\bibliography{references}

\end{document}